\documentclass[11pt]{article}
\usepackage{graphicx}
\usepackage{standalone}
\usepackage{tikz}
\usetikzlibrary{positioning}
\usetikzlibrary{backgrounds}
\usetikzlibrary{arrows}
\usepackage{latexsym}
\usepackage{amssymb}
\usepackage{amsmath}
\usepackage{amsthm}
\usepackage{forloop}
\usepackage[paper=letterpaper,margin=1in]{geometry}
\usepackage[ruled,vlined,linesnumbered]{algorithm2e}
\usepackage{nicefrac}
\usepackage{paralist}
\usepackage{hyperref}

\hypersetup{
  colorlinks   = true, 
  urlcolor     = blue, 
  linkcolor    = blue, 
  citecolor   = red 
}

\newtheorem{theorem}{Theorem}[section]
\newtheorem{lemma}[theorem]{Lemma}

\newtheorem{corollary}[theorem]{Corollary}
\newtheorem{definition}{Definition}[section]
\newtheorem{proposition}[theorem]{Proposition}
\newtheorem{observation}[theorem]{Observation}

\newtheorem{reduction}[theorem]{Reduction}

\makeatletter
\newtheorem*{rep@theorem}{\rep@title}
\newcommand{\newreptheorem}[2]{%
\newenvironment{rep#1}[1]{%
 \def\rep@title{#2 \ref{##1}}%
 \begin{rep@theorem}}%
 {\end{rep@theorem}}}
\makeatother

\newreptheorem{theorem}{Theorem}
\newreptheorem{reduction}{Reduction}

\newcommand{\refalg}[1]{Algorithm~\ref{#1}}

\newcommand{\refreduc}[1]{Reduction~\ref{#1}}
\newcommand{\reflem}[1]{Lemma~\ref{#1}}

\newcommand{\defcal}[1]{\expandafter\newcommand\csname c#1\endcsname{{\mathcal{#1}}}}
\newcommand{\defbb}[1]{\expandafter\newcommand\csname b#1\endcsname{{\mathbb{#1}}}}
\newcounter{calBbCounter}
\forLoop{1}{26}{calBbCounter}{
    \edef\letter{\Alph{calBbCounter}}
    \expandafter\defcal\letter
		\expandafter\defbb\letter
}

\newcommand{\eps}{\varepsilon}
\newcommand{\ie}{{\it i.e.}}

\newcommand{\nnR}{{\bR_{\geq 0}}}
\DeclareMathOperator{\polylog}{polylog}
\newcommand{\mwm}[0]{the problem of finding a maximum weight set subject to a $k$-extendible constraint}
\newcommand{\wmax}{{w_{\max}}}
\newcommand{\wmin}{{w_{\min}}}
\newcommand{\imax}{{i_{\max}}}
\newcommand{\imin}{{i_{\min}}}
\newcommand{\pimax}{{i'_{\max}}}
\newcommand{\pimin}{{i'_{\min}}}
\newcommand{\Greedy}{{\texttt{Greedy}}}

\begin{document}

\title{Almost Optimal Semi-streaming\\ Maximization for $k$-Extendible Systems}
\author{Moran Feldman\thanks{Dept. of Mathematics and Computer Science, Open University of Israel. E-mail: \href{mailto:moranfe@openu.ac.il}{moranfe@openu.ac.il}}
\and
Ran Haba\thanks{Dept. of Mathematics and Computer Science, Open University of Israel. E-mail: \href{ran.ronny.haba@gmail.com}{ran.ronny.haba@gmail.com}}
}
\maketitle
\begin{abstract}
In this paper we consider the problem of finding a maximum weight set subject to a $k$-extendible constraint in the data stream model. The only non-trivial algorithm known for this problem to date---to the best of our knowledge---is a semi-streaming $k^2(1 + \eps)$-approximation algorithm (Crouch and Stubbs, 2014), but semi-streaming $O(k)$-approximation algorithms are known for many restricted cases of this general problem. In this paper, we close most of this gap by presenting a semi-streaming $O(k \log k)$-approximation algorithm for the general problem, which is almost the best possible even in the offline setting (Feldman et al., 2017).

\medskip
\noindent \textbf{Keywords:} $k$-extendible systems, streaming, combinatorial optimization, greedy algorithms
\end{abstract}
\pagenumbering{Alph}
\thispagestyle{empty}
\clearpage
\setcounter{page}{1}
\pagenumbering{arabic}

\section{Introduction} \label{sec:introduction}

Many problems in combinatorial optimization can be cast as special cases of the following general task. Given a ground set $\cN$ of weighted elements, find a maximum weight subset of $\cN$ obeying some constraint $\cC$. In general, one cannot get any reasonable approximation ratio for this general task since it captures many hard problems such as maximum independent set in graphs. However, the existing literature includes many interesting classes of constraints for which the above task becomes more tractable. In particular, in the 1970's Jenkyns~\cite{J76} and Korte and Hausmann~\cite{KH78} suggested, independently, a class of constraints named \emph{$k$-set system} constraints which represents a sweet spot between generality and tractability. On the one hand, finding a maximum weight set subject to a $k$-set system constraint captures many well known problems such as matching in hypergraphs, matroid intersection and asymmetric travelling salesperson. On the other hand, $k$-set system constraints have enough structure to allow a simple greedy algorithm to find a maximum weight set subject to such a constraint up to an approximation ratio of $k$.\footnote{$k$ is a parameter of the constraint which intuitively captures its complexity. The exact definition of $k$ is given in Section~\ref{sec:preliminaries}, but we note here that in many cases of interest $k$ is quite small. For example, matroid intersection is a $2$-set system.}

The $k$-approximation obtained by the greedy algorithm for finding a maximum weight set subject to a $k$-set system constraint was recently shown to be the best possible~\cite{BV14}. Nevertheless, over the years many works improved over it either by achieving a better guarantee for more restricted classes of constraints~\cite{FNSW11,LSV10,LSV13}, or by extending the guarantee to more general objectives (such as maximizing a submodular function)~\cite{FHK17,FNSW11,FNW78,GRST10,LMNS10,LSV10,MBK16,W12}. Unfortunately, many of the above mentioned improvements are based on quite slow algorithms. Moreover, as modern applications require the processing of increasingly large amounts of data, even the simple greedy algorithm is often viewed these days as too slow for practical use. This state of affairs has motivated recent works aiming to study the problem of finding a maximum weight set subject to a $k$-set system constraint in a Big Data oriented setting such as Map-Reduce and the data stream model. For the Map-Reduce setting, Ponte Barbosa et al.~\cite{BENW16} essentially solved this problem by presenting a $(k + O(\eps))$-approximation Map-Reduce algorithm for it using $O(1/\eps)$ rounds, which almost matches the optimal approximation ratio in the sequential setting. In contrast, the situation for the data stream model is currently much more involved.

The only non-trivial data stream algorithm known to date (as far as we know) for finding a maximum weight set subject to a general $k$-set system constraint is a $k^2(1 + \eps)$-approximation semi-streaming algorithm by Crouch and Stubbs~\cite{CS14}. As one can observe, there is a large gap between the last approximation ratio and the $k$-approximation that can be achieved in the offline setting. Several works partially addressed this gap by providing an $O(k)$-approximation semi-streaming algorithms for more restricted classes of constraints, the most general of which is known as \emph{$k$-matchoid} constraints~\cite{CK15,CGQ15,FKK18,MJK18}. However, these results cannot be considered a satisfactory solution for the gap because $k$-matchoid constraints are much less general than $k$-set system constraints.\footnote{We do not formally define $k$-matchoid constraints in this paper, but it should be noted that they usually fail to capture knapsack like constraints. For example, a single knapsack constraint in which the ratio between the largest and smallest item sizes is at most $k$ is a $k$-set system constraint, but usually not a $k$-matchoid constraint.}

In this paper we make a large step towards resolving the above gap. Specifically, we present an $\tilde{O}(k)$-approximation semi-streaming algorithm for finding a maximum weight set subject to a class of constraints, known as \emph{$k$-extendible} constraints, that was introduced by~\cite{M06} and captures (to the best of our knowledge) all the special cases of $k$-set system constraints studied in the literature to date (including, in particular, $k$-matchoid constraints). Formally, we prove the following theorem.
\begin{theorem} \label{thm:streaming_k_extendible}
There is a polynomial time semi-streaming algorithm achieving $O(k \log k)$-approx\-imation for the problem of finding a maximum weight set subject to a $k$-extendible constraint. Assuming it takes constant space to store a single element and a single weight, the space complexity of the algorithm is $O(\rho (\log k + \log \rho))$, where $\rho$ is the maximum size of a feasible set according to the constraint.
\end{theorem}
As the class of $k$-extendible constraints captures every other restricted class of $k$-set system constraints from the literature, we believe Theorem~\ref{thm:streaming_k_extendible} represents the final intermediate step before closing the above mentioned gap completely (\ie, either finding an $\tilde{O}(k)$ semi-streaming algorithm for $k$-set system constraints, or proving that this cannot be done). It should also be mentioned that the approximation ratio guaranteed by Theorem~\ref{thm:streaming_k_extendible} is optimal up to an $O(\log k)$ factor since it is known that one cannot achieve better than $k$-approximation for finding a maximum weight set subject to a $k$-extendible constraint even in the offline setting~\cite{FHK17}.

\subsection{Additional Related Work}

In the $k$-dimensional matching problem, one is given a weighted hypergraph in which the vertices are partitioned into $k$ subsets, and every edge contains exactly one vertex from each one of these subsets. The objective in this problem is to find a maximum weight matching in the hypergraph. Hazan et al.~\cite{HSS06} showed that no algorithm can achieve a better than $\Omega(k / \log k)$-approximation for $k$-dimension matching unless $\mathtt{P} = \mathtt{NP}$. Interestingly, it turns out that $k$-dimensional matching is captured by all the standard restricted cases of the the problem of finding a maximum weight set subject to $k$-set system constraint, and thus, the inapproximability of Hazan et al.~\cite{HSS06} extends to them as well. For most of these restricted cases this is the strongest inapproximability known, although a tight inapproximability of $k$ was proved for $k$-set system and $k$-extendible constraints by~\cite{BV14} and~\cite{FHK17}, respectively.

Complementing the hardness result of~\cite{HSS06}, some works presented algorithmic results for either $k$-dimensional matching or natural generalizations of it such as $k$-set packing~\cite{B00,HS89,SW13}.
\section{Preliminaries and Notation} \label{sec:preliminaries}

In this section we formally define some of the terms used in Section~\ref{sec:introduction} and the notation that we use in the rest of this paper. Given a ground set $\cN$, an \emph{independence system} over this ground set is a pair $(\cN, \cI)$ in which $\cI$ is a non-empty collection of subsets of $\cN$ (formally, $\varnothing \neq \cI \subseteq 2^\cN$) which is \emph{down-closed} (\ie, if $T$ is a set in $\cI$ and $S$ is a subset of $T$, then $S$ also belongs to $\cI$). One easy way to get an example of an independence system is to take an arbitrary vector space $W$, designate the set of vectors in this space as the ground set $\cN$, and make $\cI$ the collection of all independent sets of vectors in $W$. Since removing a vector from an independent set of vectors cannot make the set dependent, the pair $(\cN, \cI)$ obtained from $W$ in this way is indeed an independence system.

The above example for getting an independence system from a vector space was one of the original motivations for the study of independence systems, and thus, a lot of the terminology used for independence systems is borrowed from the world of vector spaces. In particular, a set is called \emph{independent} in a given independence system $(\cN, \cI)$ if and only if it belongs to $\cI$, and it is called a \emph{base} of the independence system if it is an inclusion-wise maximal independent set. Using this terminology, we can now define $k$-set systems.
\begin{definition}
An independence system $(\cN, \cI)$ is a $k$-set system for an integer $k \geq 1$ if for every set $S \subseteq \cN$, all the bases of $(S, 2^S \cap \cI)$ have the same size up to a factor of $k$ (in other words, the ratio between the sizes of the largest and smallest bases of $(S, 2^S \cap \cI)$ is at most $k$).
\end{definition}

An immediate consequence of the definition of $k$-set systems is that any base of such a system is a maximum size independent set up to an approximation ratio of $k$. Thus, one can get a $k$-approximation for the problem of finding a maximum size independent set in a given $k$-set system $(\cN, \cI)$ by outputting an arbitrary base of the $k$-set system, which can be done using the following simple strategy, which we call the \emph{unweighted greedy algorithm}. Start with the empty solution, and consider the elements of the ground set $\cN$ in an arbitrary order. When considering an element, add it to the current solution, unless this will make the solution dependent (\ie, not independent).

A \emph{$k$-set system constraint} is a constraint defined by a $k$-set system, and a set $S$ obeys this constraint if and only if it is independent in that $k$-set system. Note that using this notion we can refer to the problem studied in the previous paragraph as finding a maximum cardinality set subject to a $k$-set system constraint. More generally, given a weight function $w \colon \cN \to \nnR$ and a $k$-set system $(\cN, \cI)$ over the same ground set, it is often useful to consider the problem of finding a maximum weight set $S \subseteq \cN$ subject to the constraint corresponding to this $k$-set system (the weight of a set $S$ is defined as $\sum_{u \in S} w(u)$). Jenkyns~\cite{J76} and Korte and Hausmann~\cite{KH78} showed that one can get a $k$-approximation for this problem using an algorithm, known simply as the \emph{greedy algorithm}, which is a variant of the unweighted greedy algorithm that considers the elements of $\cN$ in a non-decreasing weight order. 

The definition of $k$-set systems is very general, which occasionally does not allow them to capture all the necessary structure of a given application. Thus, various stronger kinds of independent set systems have been considered over the years, the most well known of which is the intersection of $k$ matroids (which is equivalent to a $k$-set system for $k = 1$, and represents a strictly smaller class of independence systems for larger values of $k$). In this work we consider another kind of independence systems, which was originally defined by~\cite{M06}. In this definition we use the expression $S + u$ to denote the union $S \cup \{u\}$. We use the plus sign in a similar way throughout the rest of this paper.
\begin{definition}
An independence system $(\cN, \cI)$ is a $k$-extendible system for an integer $k \geq 1$ if for any two sets $S \subseteq T \subseteq \cN$ and an element $u \not \in T$ such that $S + u \in \cI$, there is a subset $Y \subseteq T \setminus S$ of size at most $k$ such that $T \setminus Y + u \in \cI$.
\end{definition}
The class of $k$-extendible systems is general enough to capture the intersection of $k$ matroids and every other restricted class of $k$-set systems from the literature that we are aware of. In contrast, it is not difficult to verify that any $k$-extendible system is a $k$-set system. Thus, the greedy algorithm provides $k$-approximation for the problem of finding a maximum weight set subject to a $k$-extendible constraint---\ie, a constraint defined by a $k$-extendible system and allowing only sets that are independent in this system.

In the data stream model version of the above problem, the elements of the ground set of a $k$-extendible system $(\cN, \cI)$ arrive one after the other in an adversarially chosen order. An algorithm for this model views the elements of $\cN$ as they arrive, and it gets to know the weight $w(u)$ of every element $u$ upon its arrival. Additionally, as is standard in the field, we assume the algorithm has access to an \emph{independence oracle} that given a set $S \subseteq \cN$ answers whether $S$ is independent. The objective of the algorithm is to output a maximum weight independent set of the $k$-extendible system. If the algorithm is allowed enough memory to store the entire input, then the data stream model version becomes equivalent to the offline version of the problem. Thus, an algorithm for this model is interesting only if it has a low space complexity. Since any algorithm for this model must use at least the space necessary for storing its output, most works on this model look for \emph{semi-streaming} algorithms, which are data stream algorithms whose space complexity is upper bounded by $O(\rho \cdot \polylog n)$---where $\rho$ is the maximum size of an independent set and $n$ is the size of the ground set. In particular, we note that the space complexity guaranteed by Theorem~\ref{thm:streaming_k_extendible} falls within this regime because $\rho \leq n$ by definition, and one can assume that $k \leq n$ because any independence system is $n$-extendible.

One can observe that the unweighted greedy algorithm (unlike the greedy algorithm itself) can be implemented as a semi-streaming algorithm because it considers the elements in an arbitrary order. This observation is crucial for our result since the algorithm we develop is heavily based on using the unweighted greedy algorithm as a subroutine (a similar use of the unweighted greedy algorithm is done by the current state-of-the-art algorithm for the problem due to Crouch and Stubbs~\cite{CS14}).

\paragraph{Paper Organization:} In Section~\ref{sec:reduction} we present a reduction that allows us to assume that the weights of the elements are powers of $k$, at the cost of losing a factor of $O(\log k)$ in the space complexity of the algorithm. Using this reduction, we present a basic version of our algorithm in Section~\ref{sec:algorithm}. This basic version presents our main new ideas, but achieves semi-streaming space complexity only under the simplifying assumption that the ratio between the maximum and minimum element weights is polynomially bounded. This simplifying assumption can be dropped using standard techniques, and we defer the details to Appendix~\ref{app:general_weights}.
\section{Reduction to \texorpdfstring{$k$}{k}-Power Weights} \label{sec:reduction}

In this section we present a reduction that allows us to assume that the weights of all the elements in the ground set $\cN$ are powers of $k$. This reduction simplifies the algorithms we present later in this paper. However, before presenting the reduction itself, let us note that we assume from this point on that $k = 2^i$ for some integer $i \geq 1$. This assumption is without loss of generality because if $k$ does not obey it, then we can increase its value to the nearest integer that does obey it. Since the new value of $k$ is larger than the old value by at most a factor of $2$, the approximation ratio guaranteed for both values of $k$ by Theorem~\ref{thm:streaming_k_extendible} is asymptotically equal.

We say that an instance of {\mwm} is a \emph{$k$-power} instance if the weights of all the elements in it are powers of $k$.
\begin{reduction} \label{reduction:power_k_weights} 
Assume that we are given a polynomial time data stream algorithm $ALG$ for \mwm{}. If $ALG$ provides $\alpha$-approximation for $k$-power instances of the problem using $S_{ALG}$ space, then there exists a  polynomial time data stream algorithm for the same problem which achieves $O(\alpha \log{k})$-approximation for arbitrary  instances using $O(S_{ALG} \cdot \log k)$ space. Moreover, if the weights of all the elements fall within some range $[\wmin, \wmax]$, then it suffices for $ALG$ to provide $\alpha$-approximation for $k$-power instances in which all the weights fall within the range $[\wmin/k, \wmax]$.
\end{reduction}

Before presenting the algorithm we use to prove the above reduction, we need to define some additional notation.
Let $\ell \triangleq \log_2 k$, and note that $\ell$ is a positive integer because we assume that $k$ is at least $2$ and a power of $2$.
For every element $u \in \cN$ of weight $w(u)$, we define an auxiliary weight $w_2(u) \triangleq k^{\lfloor \log_k w(u) \rfloor}$.
Intuitively, $w_2(u)$ is the highest power of $k$ which is not larger than $w(u)$. The following observation formally states the properties of $w_2$ that we need.
In this observation we use the notation $i(u) \triangleq \lfloor \log_2 w(u) \rfloor$. 

\begin{observation} \label{obs:weights_relationships}
For every element $u \in \cN$, $w_2(u)$ is a power of $k$ obeying $w(u)/2 \leq w_2(u) \cdot 2^{i(u) \bmod \ell} \leq w(u)$ and $w(u) / k \leq w_2(u) \leq w(u)$.
\end{observation}

\begin{proof}
The first part of the observation, namely that $w_2(u)$ is a power of $k$, follows immediately from the definition of $w_2$. 
Thus, we concentrate here on proving the other parts of the observation.

Note that
\[
	w_2(u)
	=
	k^{\lfloor \log_k w(u) \rfloor}
	=
	k^{\lfloor \ell^{-1} \cdot \log_2 w(u)\rfloor}
	=
	k^{\ell^{-1} \cdot \{\lfloor \log_2 w(u)\rfloor - \lfloor \log_2 w(u) \rfloor \bmod \ell\}}
	=
	k^{\ell^{-1} \cdot \lfloor \log_2 w(u)\rfloor} / 2^{i(u) \bmod \ell}
	\enspace.
\]
Rearranging the last equality, we get
\[
	\frac{w(u)}{2}
	=
	k^{\log_k w(u) - \log_k 2}
	=
	k^{\ell^{-1} \log_2 w(u) - \ell^{-1}}
	\leq
	k^{\ell^{-1} \cdot \lfloor \log_2 w(u) \rfloor}
	=
	w_2(u) \cdot 2^{i(u) \bmod \ell}
	\enspace,
\]
and
\[
	w_2(u) \cdot 2^{i(u) \bmod \ell}
	=
	k^{\ell^{-1} \cdot \lfloor \log_2 w(u) \rfloor}
	\leq
	k^{\ell^{-1} \cdot \log_2 w(u)}
	=
	k^{\log_k w(u)}
	=
	w(u)
	\enspace.
\]

To complete the proof of the observation, we note that it also holds that
\[
	w_2(u)
	=
	k^{\lfloor \log_k w(u) \rfloor}
	\leq
	k^{\log_k w(u)}
	=
	w(u)
	\quad
	\text{and}
	\quad
	w_2(u)
	=
	k^{\lfloor \log_k w(u) \rfloor}
	\geq
	k^{\log_k w(u) - 1}
	=
	\frac{w(u)}{k}
	\enspace.
	\qedhere
\]
\end{proof}

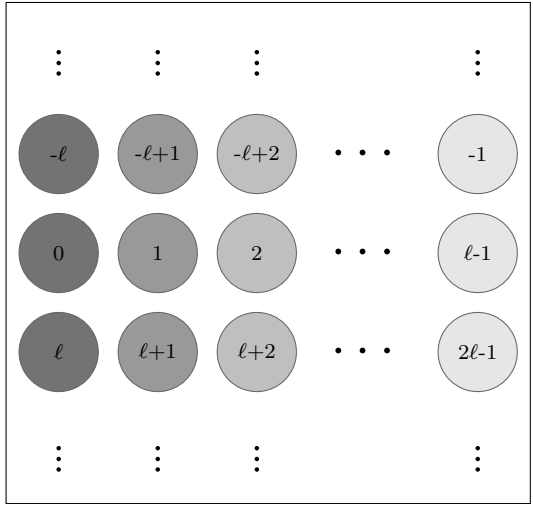
\begin{figure}[ht]
	\centering
 	\definecolor{red}{gray}{0.9}
\definecolor{yellow}{gray}{0.75}
\definecolor{orange}{gray}{0.6}
\definecolor{green}{gray}{0.45}

\tikzstyle{every node}=[font=\scriptsize]

\begin{tikzpicture}[
framed,
align=center,
node distance=0.25cm,
roundnode/.style={circle, draw=black!60, minimum width={width("-2L+2")},},
blanknode/.style={circle, draw=white!60, minimum width={width("-2L+2")},font=\huge},
]

\node[blanknode]					(-2L)                       			{$\vdots$};
\node[blanknode]     			  	(-2L+1)       	[right=of -2L] 			{$\vdots$};
\node[blanknode]     			  	(-2L+2)       	[right=of -2L+1] 		{$\vdots$};
\node[blanknode]			       	(q_dots0)      	[right=of -2L+2] 		{};

\node[roundnode, fill=green]		(-L)			[below=of -2L]			{-$\ell$};
\node[roundnode, fill=orange]       (-L+1)       	[right=of -L] 			{-$\ell$+1};
\node[roundnode, fill=yellow]       (-L+2)       	[right=of -L+1] 		{-$\ell$+2};
\node[blanknode]			       	(q_dots1)      	[right=of -L+2] 		{$\cdots$};
\node[roundnode, fill=red]     		(-1)       		[right=of q_dots1] 		{-1};

\node[roundnode, fill=green]       	(0)				[below=of -L]			{0};
\node[roundnode, fill=orange]       (1)       		[right=of 0] 			{1};
\node[roundnode, fill=yellow]       (2)		       	[right=of 1]	 		{2};
\node[blanknode]			       	(q_dots2)      	[right=of 2] 			{$\cdots$};
\node[roundnode, fill=red]       	(L-1)      		[right=of q_dots2] 		{$\ell$-1};

\node[roundnode, fill=green]       	(L)				[below=of 0]			{$\ell$};
\node[roundnode, fill=orange]       (L+1)       	[right=of L] 			{$\ell$+1};
\node[roundnode, fill=yellow]       (L+2)		    [right=of L+1]	 		{$\ell$+2};
\node[blanknode]			       	(q_dots3)      	[right=of L+2] 			{$\cdots$};
\node[roundnode, fill=red]       	(2L-1)      	[right=of q_dots3] 		{2$\ell$-1};

\node[blanknode]					(2L)            [below=of L]           	{$\vdots$};
\node[blanknode]       				(2L+1)       	[right=of 2L] 			{$\vdots$};
\node[blanknode]       				(2L+2)       	[right=of 2L+1] 		{$\vdots$};
\node[blanknode]       				(q_dots4)       [right=of 2L+2] 		{};

\node[blanknode]     			 	(-L-1)       	[above=of -1] 		{$\vdots$};
\node[blanknode]      				(3L-1)       	[below=of 2L-1] 		{$\vdots$};


\end{tikzpicture}
 	
 	\caption{Each circle in this drawing represent a class, and the value $i(u)$ of the elements in this class appears in the center the circle. The classes are grouped according to the columns in the drawing. We note that element $u$ belonging to group $j$ has weight within the range $[2^{k \ell +j}, 2^{k \ell +j + 1})$, where $k$ is an integer and $i(u)= k \ell +j$.}	\label{fig}
\end{figure}

We are now ready to present the algorithm that we use to prove \refreduc{reduction:power_k_weights}, which appears as \refalg{alg}.
To intuitively understand this algorithm, it is useful to think of $i(u)$ as the ``class'' element $u$ belongs to. All the elements within class $i$ have weights between $2^i$ and $2^{i + 1}$, and thus, treating them all as having the weight $2^i$ does not affect the approximation ratio by more than a factor of $2$. Let us call $2^i$ the \emph{characteristic} weight of class $i$. Note now that the ratio between the characteristic weight of class $i_1$ and the characteristic weight of class $i_2$ is $2^{i_1 - i_2}$, which is a power of $k$ whenever $i_1 - i_2$ is an integer multiple of $\ell = \log_2 k$. Thus, one can group the classes into $\ell$ groups such that the ratio between the characteristic weights of any pair of classes within a group is a power of $k$ (see Figure~\ref{fig} for a graphial illustration of these groups). Moreover, by multiplying all the characteristic weights in the group by an appropriate scaling factor, one can make them all powers of $k$. This means that for every group there exists a transformation that converts all the weights of the elements in it to powers of $k$ and preserves the ratio between any two weights in the group up to a factor of $2$. In particular, we get that the elements of the group after the transformation form a $k$-power instance.

Adding up all the above, we have described a way to transform any instance of finding a maximum weight independent set subject to a $k$-extendible constraint into $\ell$ new instances of this problem that are guaranteed to be $k$-power. Algorithm~\ref{alg} essentially creates these $\ell$ new instances on the fly, and feeds them to $\ell$ copies of the algorithm $ALG$ whose existence is assumed in Reduction~\ref{reduction:power_k_weights}. Given this point of view, $i(u) \bmod \ell$ should be understood as the group to which element $u$ belongs, and $w_2(u)$ is the transformed weight of $u$. Observation~\ref{obs:weights_relationships} can now be interpreted as stating that the ratio between the weights of elements belonging to the same group (and thus, having the same $i(u) \bmod \ell$ value) is indeed changed by the transformation by at most a factor of $2$.

\begin{algorithm} \label{alg}
\DontPrintSemicolon
\caption{\textbf{Modulo $\ell$ Split}}
Create $\ell$ instances of $ALG$ named $ALG_0, ALG_1, \dotsc, ALG_{\ell - 1}$.\\
\For{each element $u$ that arrives from the stream}
{
	Calculate $i(u)$ and $w_2(u)$ as defined above. \\
	Feed $u$ to $ALG_{(i(u) \bmod \ell)}$ with the weight $w_2(u)$. \\
}	
Let $C_i$ denote the output of $ALG_i$ for every $0 \leq i \leq \ell - 1$. \\
\Return{the best solution among $C_0, C_1, \dotsc, C_{\ell - 1}$}.
\end{algorithm}

In the rest of this section, we use $B_i$ to denote the set of elements fed to instance $ALG_i$ by \refalg{alg}, and $T$ to denote the output of \refalg{alg}.
Additionally, we denote by $OPT$ an arbitrary (fixed) optimal solution for the original instance recieved by Algorithm~\ref{alg}.
The following lemma proves that \refalg{alg} has the approximation ratio guaranteed by \refreduc{reduction:power_k_weights}.

\begin{lemma} \label{lemma:logk_approx}
$w(OPT) \leq O(\alpha \log k) \cdot w(T)$.
\end{lemma}
\begin{proof}
Since \refalg{alg} feeds every arriving element into exactly one of the instances $ALG_0,\allowbreak ALG_1, \dotsc, ALG_{\ell-1}$, the sets $B_0, B_1, \dotsc, B_{\ell-1}$ form a disjoint partition of $\cN$. Thus,
\[
	w(OPT) = \sum \limits_{i=0}^{\ell-1} w(B_i \cap OPT)
	\enspace .
\]
Hence, by an averaging argument, there must exist an index $i$ such that $w(OPT) \leq \ell \cdot w(OPT \cap B_i) $.

We now note that it follows from the pseudocode of Algorithm~\ref{alg} and Observation~\ref{obs:weights_relationships} that the copies of $ALG$ get only weights that are powers of $k$, and moreover, these weights belong to the range $[\wmin/k, \wmax]$ whenever the original weights received by Algorithm~\ref{alg} belong to the range $[\wmin, \wmax]$. Thus, by the assumption of Reduction~\ref{reduction:power_k_weights}, $ALG_i$ achieves $\alpha$-approximation for the instance it faces. Since $B_i \cap OPT$ is a feasible solution within this instance and $C_i$ is the output of $ALG_i$, we get $w_2(OPT \cap B_i) \leq \alpha \cdot w_2(C_i)$.
Therefore,
\[
	w(OPT)
	\leq
	\ell \cdot w(OPT \cap B_i)
	\leq
	2\ell \cdot w_2(OPT \cap B_i) \cdot 2^i
	\leq
	2\ell\alpha \cdot w_2(C_i) \cdot 2^i
	\leq
	2\ell\alpha \cdot w(C_i)
	\leq
	2\ell\alpha \cdot w(T)
\enspace,
\]
where the second and penultimate inequalities hold by Observation~\ref{obs:weights_relationships}, and the last inequality is due to the fact that $T$ is the best solution among $C_0, C_1, \dotsc, C_{\ell-1}$.
\end{proof}

The next lemma analyzes the space complexity of Algorithm~\ref{alg} and completes the proof of \refreduc{reduction:power_k_weights}.

\begin{lemma}
\refalg{alg}'s space complexity is $O(S_{ALG} \cdot \log{k})$.
\end{lemma}
\begin{proof}
\refalg{alg} runs $\log{k}$ parallel copies of $ALG$, each of them is assumed (by \refreduc{reduction:power_k_weights}) to use $S_{ALG}$ space. Thus, the space required by these $\log k$ copies is $O(S_{ALG} \cdot \log k)$. In addition to this space, Algorithm~\ref{alg} only requires enough space to do two things.
\begin{itemize}
	\item The algorithm has to store the outputs of the copies of $ALG$. However, these outputs are originally stored by the copies themselves, and thus, storing them requires no more space than what is used by the copies.
	\item Calculate the sum of the weights of the elements in the solutions produced by the copies of $ALG$. 
	Since we assume that the weight of an element can be stored in constant space, this requires again (up to constant factors) no more space than the space used by the copies of $ALG$ to store their solutions. \qedhere
\end{itemize}
\end{proof}

\section{Algorithm} \label{sec:algorithm}

In this section we present a data stream algorithm for $k$-power instances of {\mwm}. This algorithm assumes access to positive upper bound $\wmax$ and lower bound $\wmin$ on the weights of all the elements, and has a semi-streaming space complexity when the ratio between $\wmax$ and $\wmin$ is upper bounded by a polynomial in $n$. Proposition~\ref{prop:k_approx_for_k_power} states the properties that we prove for this algorithm more formally.

\begin{proposition} \label{prop:k_approx_for_k_power}
There exists a $2k$-approximation data stream algorithm for \emph{$k$-power} instances of {\mwm}. This algorithm assumes access to positive upper bound $\wmax$ and lower bound $\wmin$ on the weights of all the elements, and its space complexity is $O(\rho (\log(\nicefrac{\wmax}{\wmin}) / \log k + 1))$ under the assumption that constant space suffices to store a single element and a single weight.
\end{proposition}

Before getting to the proof of Proposition~\ref{prop:k_approx_for_k_power}, we note that together with Reduction~\ref{reduction:power_k_weights} this proposition immediately implies the following corollary.
\begin{corollary} \label{cor:theorem_for_bounded_weights}
There exists an $O(k \log k)$-approximation data streaming algorithm for {\mwm}. This algorithm assumes access to positive upper bound $\wmax$ and lower bound $\wmin$ on the weights of all the elements. The space complexity of this algorithm is $O(\rho (\log(\nicefrac{\wmax}{\wmin}) + \log k))$ under the assumption that constant space suffices to store a single element and a single weight.
\end{corollary}

Note that when the ratio between $\wmax$ and $\wmin$ is polynomial in $n$, the space complexity of the algorithm from Corollary~\ref{cor:theorem_for_bounded_weights} becomes $O(\rho \log n)$, and thus, the algorithm is semi-streaming. In Appendix~\ref{app:general_weights} we explain how the algorithm can be modified so that it keeps the ``effective'' ratio $\nicefrac{\wmax}{\wmin}$ on the order of $O(k^2\rho^2)$ even when no values $\wmax$ and $\wmin$ are supplied to the algorithm and the weights of the elements come from an arbitrary range. This leads to the space complexity of $O(\rho (\log k + \log \rho))$ stated in Theorem~\ref{thm:streaming_k_extendible}.

The rest of this section is devoted to the proof of Proposition~\ref{prop:k_approx_for_k_power}. As a first step towards this goal, let us recall that the unweighted greedy algorithm is an algorithm that considers the elements of the ground set $\cN$ in an arbitrary order, and adds every considered element to the solution it constructs if that does not violate independence. As mentioned above, it follows immediately from the definition of $k$-set systems that the unweighted greedy algorithm achieves an approximation ratio of $k$ for the problem of finding a maximum cardinality independent set subject to a $k$-set system constraint. Since $k$-set systems generalize $k$-extendible systems, the same is true also for $k$-extendible constraints. The following lemma improves over this by showing a tighter guarantee for $k$-extendible constraints.

\begin{lemma}
\label{lem:greedy_improved_result}
Given a $k$-extendible set system $(\cN, \cI)$, the unweighted greedy algorithm is guaranteed to produce an independent set $B$ such that $k \cdot |B \setminus A| \geq |A \setminus B|$ for any independent set $A \in \cI$.
\end{lemma}

\begin{proof}
Let us denote the elements of $B \setminus A$ by $x_1,x_2,\dotsc,x_m$ in an arbitrary order. Using these elements, we recursively define a series of independent sets $A_0, A_1,\dotsc,A_m$. The set $A_0$ is simply the set $A$. For $1 \leq i \leq m$, we define $A_i$ using $A_{i - 1}$ as follows. Since $(\cN, \cI)$ is a $k$-extendible system and the subsets $A_{i-1}$ and $A_{i-1} \cap B + x_i \subseteq B$ are both independent, there must exist a subset $Y_i \subseteq A_{i-1} \setminus (A_{i-1} \cap B) = A_{i-1} \setminus B$ such that $|Y_i| \leq k$ and $A_{i-1} \setminus Y_i+x_i\in \mathcal{I}$. Using the subset $Y_i$, we now define $A_i = A_{i-1} \setminus Y_i+x_i$.
Note that by the definition of $Y_i$, $A_i \in \mathcal{I}$ as promised.
Furthermore, since $Y_i \cap B = \varnothing$ for each $0 \leq i \leq m$, we know that $(A \cup \{x_1, x_2, \dots, x_m\}) \cap B \subseteq A_m$, which implies $B \subseteq A_m$ because $\{x_1,x_2,\dotsc,x_m\} = B \setminus A$.
However, $B$, as the output of the unweighted greedy algorithm, must be inclusion-wise maximal independent set (\ie, a base), and thus, it must be in fact equal to the independent set $A_m$ containing it.

Let us now denote $Y = \bigcup_{i=1}^{m} Y_i$, and consider two different ways to bound the number of elements in $Y$. On the one hand, since every set $Y_i$ includes up to $k$ elements, we get $|Y| \leq km = k \cdot |B \setminus A|$. On the other hand, the fact that $B = A_m$ implies that every element of $A \setminus B$ belongs to $Y_i$ for some value of $i$, and therefore, $|Y| \geq |A \setminus B|$. The lemma now follows by combining these two bounds. 
\end{proof}

We are now ready to present the algorithm we use to prove Proposition~\ref{prop:k_approx_for_k_power}, which is given as \refalg{cs_alg}.
This algorithm has two main stages. In the first stage, the algorithm runs an independent copy of the unweighted greedy algorithm for every possible weight of elements. The copy corresponding to the weight $k^i$ is denoted by $\Greedy_i$ in the pseudocode of the algorithm, and Algorithm~\ref{cs_alg} feeds to it only the input elements whose weight is at least $k^i$. The output of $\Greedy_i$ is denoted by $C_i$ in the algorithm. We also denote in the analysis by $E_i$ the set of elements fed to $\Greedy_i$. By definition, $C_i$ is obtained by running the unweighted greedy algorithm on the elements of $E_i$, which is a property we use below.

In the second stage of \refalg{cs_alg} (which is done as a post-processing after the stream has ended),  the algorithm constructs an output set $T$ based on the outputs of the copies of the unweighted greedy algorithm. Specifically, this is done by running the unweighted greedy algorithm on the elements of $\bigcup_{i = \imin}^{\imax} C_i$, considering the elements of the sets $C_i$ in a decreasing value of $i$ order. While doing so, the given pseudocode also keeps in $T_i$ the temporary solution obtained by the unweighted greedy algorithm after considering only the elements of $C_j$ for $j \geq i$. This temporary solution is used by the analysis below, but need not be kept by a real implementation of \refalg{cs_alg}.

\begin{algorithm} 
\DontPrintSemicolon
	\caption{\textbf{Greedy of Greedies}}
	\label{cs_alg}

	Let $\imin \gets \lceil \log_k \wmin \rceil$ and $\imax \gets \lfloor \log_k \wmax \rfloor$.\\
	Create 
	$\imax - \imin + 1$ instances of the unweighted greedy algorithm named $\Greedy_{\imin}, \Greedy_{\imin + 1}, \dotsc, \Greedy_{\imax}$.\\

	\For{each element $u$ that arrives from the stream}
	{
		Let $i_u \gets \log_k w(u)$.\\
		Feed $u$ to $\Greedy_{\imin},\Greedy_{\imin + 1},\dotsc,\Greedy_{i_u}$.
	}
	
	\BlankLine

	Let $C_i$ denote the output of $\Greedy_i$ for every $\imin \leq i \leq \imax$. \\

	Let $T \gets \varnothing$. \\
	\For{every $\imin \leq i \leq \imax$ in descending order}
	{
		Greedily add elements from $C_i$ to $T$ as long as this is possible. \\
		Let $T_i$ denote the current value of $T$.
	}
	\Return{$T$}.
\end{algorithm}

We begin the analysis of \refalg{cs_alg} by analyzing its space complexity.
\begin{lemma} \label{lem:space_complexity}
\refalg{cs_alg} can be implemented using a space complexity of $O(\rho (\log(\nicefrac{\wmax}{\wmin})/\log k \allowbreak + 1))$.
\end{lemma}

\begin{proof}
Note that each copy of the unweighted greedy algorithm only has to store its solution, which contains up to $\rho$ elements since it is independent. \refalg{cs_alg} uses $\imax - \imin + 1$ such copies, and thus, the space it needs for these copies is only
\[
	\rho(\imax - \imin + 1)
	\leq
	\rho \left(\log_k\left(\frac{\wmax}{\wmin}\right) + 1\right)
	=
	\rho \cdot O\left(\frac{\log(\nicefrac{\wmax}{\wmin})}{\log k} + 1\right)
	\enspace.
\]

In addition to the space used by the copies of the unweighted greedy algorithm, Algorithm~\ref{cs_alg} only needs to store the set $T$. This set contains a subset of the elements from the outputs of the above copies, and thus, can increases the space required only by a constant factor.
\end{proof}

To complete the proof of Proposition~\ref{prop:k_approx_for_k_power}, it remains to analyze the approximation ratio of Algorithm~\ref{cs_alg}. We begin with the following lemma, which is the technical heart of our analysis. Like in Section~\ref{sec:reduction}, let us denote by $OPT$ be an arbitrary (fixed) optimal solution to the problem we want to solve. We also assume for consistency that $T_{\imax + 1} = \varnothing$ (note that $T_{\imax + 1}$ is not defined by Algorithm~\ref{cs_alg}).

\begin{lemma}
\label{lem:size_of_Opt_intersection_Ei}
For each integer $\imin \leq i \leq \imax$, $k^2\cdot|T_{i+1}| + k\cdot|T_i \setminus T_{i+1}| \geq |OPT \cap E_i|$.
\end{lemma}

\begin{proof}
The set $T_i$ can be viewed as the output of the unweighted greedy algorithm running on $\bigcup_{i \leq j \leq \imax} C_j$. 
Since we also know that $C_i$ is independent, \reflem{lem:greedy_improved_result} guarantees
\[
	k \cdot |T_i \setminus C_i|
	\geq
	|C_i \setminus T_i|
	\enspace.
\]
Adding $k \cdot |C_i \cap T_i|$ to both its sides, we get
\begin{align*}
	k \cdot |T_i|
	\geq{} &
	k \cdot |C_i \cap T_i| + |C_i \setminus T_i|
	=
	k \cdot |C_i \cap T_i| + \{|C_i| - |C_i \cap T_i|\} \\
	={} &
	(k - 1) \cdot |C_i \cap T_i| + |C_i| 
	\geq
	(k - 1) \cdot |C_i \cap T_i| + k^{-1} \cdot |OPT \cap E_i|
	\enspace ,
\end{align*}
where the last inequality holds since the unweighted greedy algorithm achieves $k$-approximation and $OPT \cap E_i$ is an independent set within $E_i$ (recall that $E_i$ is the set of elements that were fed to $\Greedy_i$). Using the last inequality we can now get
\begin{align*}
k \cdot |T_i \setminus T_{i+1}| + k \cdot |T_{i+1}|
={} &
k \cdot |T_i|
\geq
(k - 1) \cdot |C_i \cap T_i| + k^{-1} \cdot |OPT \cap E_i| \\
\geq{} &
(k - 1) \cdot |T_i \setminus T_{i+1}| + k^{-1} \cdot |OPT \cap E_i|
\enspace ,
\end{align*}
where the first equality holds because $T_{i+1} \subseteq T_i$, and the second inequality holds because $T_i \setminus T_{i+1} \subseteq C_i \cap T_i$ (recall that the algorithm constructs $T_i$ by adding elements of $C_i$ to $T_{i + 1}$).
The lemma now follows by rearranging the above inequality and multiplying it by $k$.
\end{proof}

Using the last lemma, we can prove the existence of a useful mapping from the elements of $OPT$ to the elements of $T$.
\begin{lemma} \label{lem:mapping}
There exists a mapping $f\colon OPT \to T$ such that
\begin{compactenum}
\item for each $t \in T$, $|f^{-1}(t)| \leq k^2$. \label{prop:budget_large}
\item for each $t \in T$, $|\{u \in f^{-1}(t) \mid w(u) = w(t)\}| \leq k$. \label{prop:budget_small} 
\item for each $u \in OPT$, $w(u)\leq w(f(u))$.\label{prop:weights}
\end{compactenum}
\end{lemma}

\begin{proof}
We construct $f$ by scanning the elements $OPT$ and defining the mapping $f(e)$ for every element $e$ scanned.
To describe the order in which we scan the elements of $OPT$, let us define $P_i = OPT \cap (E_i \setminus E_{i - 1})$.
Note that $P_{i_{\min}}, P_{i_{\min} + 1}, \dotsc, P_{i_{\max}}$ is a disjoint partition of $OPT$, and thus, any scan of the elements of $P_{i_{\min}}, P_{i_{\min} + 1}, \dotsc, P_{i_{\max}}$ is a scan of the elements of $OPT$.
Specifically, we scan the elements of $OPT$ by first scanning the elements of $P_{i_{\max}}$ in an arbitrary order, then scanning the elements of $P_{i_{\max} - 1}$ in an arbitrary order, and so on.
Consider now the situation when our scan gets to an arbitrary element $u$ of set $P_i$.
One can note that prior to scanning $u$, we scanned (and mapped) only elements of $P_i \cup P_{i + 1} \cup \dotsb \cup P_{i_{\max}} = OPT \cap E_i$, and thus, we mapped at most $|OPT \cap E_i| - 1$ elements (the $-1$ is due to the fact that $u \in OPT \cap E_i$, and $u$ was not mapped yet).
Combining this with \reflem{lem:size_of_Opt_intersection_Ei}, we get that at the point in which we scan $u$ there must still be either an element $t \in T_{i+1}$ that still has less than $k^2$ elements mapped to it or an element $t \in T_i \setminus T_{i + 1}$ that still has less than $k$ elements mapped to it.
We choose the mapping $f(u)$ of $u$ to be an arbitrary such element $t$.

Property~\ref{prop:budget_large} of the lemma is clearly satisfied by the above construction because we never map an element $u$ to an element $t$ that already has $k^2$ elements mapped to it.
To see why Property~\ref{prop:weights} of the lemma also holds, note that every element $u \in P_i$ must have a weight of $k^i$ by the definition of $P_i$.
This element is mapped by $f$ to some element $t \in T_{i + 1} \cup (T_i \setminus T_{i + 1}) = T_i \subseteq E_i$, and the weight of $t$ is at least $k^i = w(u)$ by the definition of $E_i$.
It remains to prove Property~\ref{prop:budget_small} of the lemma. Consider an arbitrary element $t \in T$ of weight $k^i$. The elements of $OPT$ whose weight is $k^i$ are exactly the elements of $P_i$, and thus, we need to show that $|f^{-1}(t) \cap P_i| \leq k$.
Since all the elements of $T_{i + 1} \subseteq C_{i + 1} \cup C_{i + 2} \cup \dotsb \cup C_{i_{\max}} \subseteq E_{i + 1}$ have weights of at least $k^{i + 1}$, $t$ cannot belong to $T_{i + 1}$.
Thus, an element of $P_i$ can be mapped to $t$ when scanned only if $t$ has less than $k$ elements already mapped to it (if $t \in T_i$) or not at all (if $t \not \in T_i$), which implies that no more than $k$ elements of $P_i$ can get mapped to $t$, which is exactly what we wanted to prove.
\end{proof}

We are now ready to prove the approximation ratio of Algorithm~\ref{cs_alg} (and complete the proof of Proposition~\ref{prop:k_approx_for_k_power}).

\begin{lemma} \label{lem:approx_ratio_cs_alg}
\refalg{cs_alg} is a $2k$-approximation algorithm for $k$-power instances of {\mwm}.
\end{lemma}
\begin{proof}
Let $f$ be the function whose existence is guaranteed by \reflem{lem:mapping}.
The properties of this function imply that, for each element $t\in T$,
\[
\sum_{u \in f^{-1}(t)} \mspace{-9mu} w(u)
=
\sum_{ \substack{ u \in f^{-1}(t) \\ w(u) 
= 
w(t)}} \mspace{-9mu} w(u) +  \sum_{ \substack{e \in f^{-1}(t) \\ w(u) < w(t)}} \mspace{-9mu} w(u) 
\leq 
k \cdot w(t) + (k^2-k) \cdot \frac{w(t)}{k} \leq 
2 k  \cdot w(t)
\enspace .
\]
Thus,
\[
w(OPT) 
=
 \sum_{u \in OPT}  \mspace{-9mu}w(u) 
= 
\sum_{t \in T} \sum_{u \in f^{-1}(t)} \mspace{-9mu} w(u) 
\leq 
\sum_{t \in T} \mspace{9mu} [2  k  \cdot w(t)] 
= 
2  k \cdot w(T)
\enspace ,
\]
which completes the proof of the lemma.
\end{proof}
\section{Conclusion}

In this work we have presented the first semi-streaming $\tilde{O}(k)$-approximation algorithm for the problem of finding a maximum weight set subject to a $k$-extendible constraint. This result is intrinsically interesting because the generality of $k$-extendible constraints makes our algorithm applicable to many problems of interest. Additionally, we believe (as discussed in Section~\ref{sec:introduction}) that our result is likely to be the final intermediate step towards the goal of designing an algorithm with similar properties for general $k$-set system constraints or proving that this cannot be done.

Given our work, the immediate open question is to settle the approximation ratio that can be obtained for $k$-set system constraints in the data stream model. Another interesting research direction is to find out whether one can improve over the approximation ratio of our algorithm. Specifically, we leave open the question of whether there is a semi-streaming algorithm for finding a maximum weight set subject to a $k$-extendible constraint whose approximation ratio is clean $O(k)$.

\bibliographystyle{plain}
\bibliography{kExtendible}

\appendix
\section{Algorithm for General Weights} \label{app:general_weights}

In this section we present a semi-streaming algorithm for $k$-power instances of {\mwm}. Unlike Algorithm~\ref{cs_alg}, this algorithm does not assume access to the bounds $\wmax$ and $\wmin$, and its space complexity remains nearly linear regardless of the ratio between these bounds. A more formal statement of the properties of this algorithm is given in Proposition~\ref{prop:k_approx_for_k_power}. Note that, together with Reduction~\ref{reduction:power_k_weights}, this proposition immediately implies Theorem~\ref{thm:streaming_k_extendible}.
\begin{proposition} \label{prop:k_approx_for_k_power_general}
There exists a $4k$-approximation semi-streaming algorithm for \emph{$k$-power} instances of {\mwm}. The space complexity of this algorithm is $O(\rho (\log k + \log \rho) / \log k)$ under the assumption that constant space suffices to store a single element and a single weight.
\end{proposition}

Throughout this section we assume for simplicity that the $k$-extendible system does not include any self-loops (a \emph{self-loop} is an element $u \in \cN$ such that $\{u\}$ is a dependent set---\ie, $\{u\} \not \in \cI$). This assumption is without loss of generality since a self-loop cannot belong to any independent set, and thus, an algorithm can safely ignore self-loops if they happen to exist. One consequence of this assumption is that $\max_{u \in \cN} w(u) \leq w(OPT)$, where $OPT$ is an arbitrary fixed optimal solution like in the previous sections. This inequality holds since $\{u\}$ is a feasible solution for every element $u \in \cN$, and therefore, its weight cannot exceed the weight of $OPT$.

As mentioned in Section~\ref{sec:algorithm}, the algorithm we use to prove Proposition~\ref{prop:k_approx_for_k_power_general} is a variant of \refalg{cs_alg} that includes additional logic designed to force the ratio $\nicefrac{\wmax}{\wmin}$ to be effectively polynomial---specifically, $O(k^2\rho^2)$. Given access to $\rho$ and $\max_{u \in \cN} w(u)$, this could be done simply by settings $\wmax = \max_{u \in \cN} w(u)$ and $\wmin = \max_{u \in \cN} w(u) / (2\rho)$ and discarding any element whose weight is lower then $\wmin$.\footnote{Starting from this point, $\wmax$ and $\wmin$ are no longer necessarily upper and lower bounds on the weights of all the elements. However, they remain upper and lower bounds on the weights of the non-discarded elements.} This guarantees that the ratio $\nicefrac{\wmax}{\wmin}$ is small, and affects the weight of the optimal solution $OPT$ by at most a constant factor since the total weight of the elements of this solution that get discarded is upper bounded by
\[
	|OPT| \cdot \wmin
	\leq
	\rho \cdot \frac{\max_{u \in \cN} w(u)}{2\rho}
	=
	\frac{\max_{u \in \cN}w(u)}{2}
	\leq
	\frac{w(OPT)}{2}
	\enspace.
\]

Unfortunately, our algorithm does not have access (from the beginning) to $\rho$ and $\max_{u \in \cN} w(u)$. As an alternative, this algorithm, which is given as Algorithm~\ref{algorithm:general_weights}, does two things. First, it keeps $\wmax$ equal to the maximum weight of the elements seen so far, which guarantees that eventually $\wmax$ becomes $\max_{u \in \cN} w(u)$. Second, it runs the unweighted greedy algorithm on the input it receives. The size of the solution maintained by the unweighted greedy algorithm, which we denoted by $g$, provides an estimate for the maximum size of an independent set consisting only of elements that have already arrived. In particular, after all the elements arrive, $\rho/k \leq g \leq \rho$ because the unweighted greedy algorithm is a $k$-approximation algorithm.

Given the above discussion and the fact that the final value of $kg$ is an upper bound on $\rho$, it is natural to define $\wmin$ as $\wmax / (2kg)$ and discard every element whose weight is lower than $\wmin$. Unfortunately, this does not work since $\wmax$ and $g$ change during the execution of Algorithm~\ref{algorithm:general_weights}, and reach their final values only when it terminates. Thus, we need to set $\wmin$ to a more conservative (lower) value. In particular, Algorithm~\ref{algorithm:general_weights} uses $\wmin = \wmax / (2gk)^2$.

Like Algorithm~\ref{cs_alg}, Algorithm~\ref{algorithm:general_weights} maintains an instance of the unweighted greedy algorithm for every possible weight between $\wmin$ and $\wmax$. However, doing so is somewhat more involved for Algorithm~\ref{algorithm:general_weights} because $\wmin$ and $\wmax$ change during the algorithm's execution, which requires the algorithm to occasionally create and remove instances of unweighted greedy. The creation of such instances involves one subtle issue that needs to be kept in mind. In Algorithm~\ref{cs_alg} every instance of unweighted greedy associated with a weight $w$ receives all elements whose weight is at least $w$. To mimic this behavior, when Algorithm~\ref{algorithm:general_weights} creates new instances of unweighted greedy following a decrease in $\wmin$ (which can happen when $g$ increases), the newly created instances are not fresh new instances but copies of the instance of unweighted greedy that was previously associated with the lowest weight.

The rest of the details of Algorithm~\ref{algorithm:general_weights} are identical to the details of Algorithm~\ref{cs_alg}. Specifically, every arriving element $u$ is feed to every instance of unweighted greedy associated with a weight of $w(u)$ or less, and at termination the outputs of all the unweighted greedy instances are combined in the same way in which this is done in Algorithm~\ref{cs_alg}.

\begin{algorithm}[ht]
\DontPrintSemicolon
	\caption{\textbf{Greedy of Greedies for Unbounded Weights}}
	
\label{algorithm:general_weights}

	Create an instance of the unweighted greedy algorithm named $\Greedy$, and let $g$ denote the size of the solution maintained by it. \\
	
	\For{each element $u$ that arrives from the stream}
	{
		Feed $u$ to $\Greedy$. \\
		\If{$u$ is the first element to arrive}
		{
			Let $\wmax \gets w(u)$ and $\wmin \gets \wmax / (2gk)^2$.\\
			Let $\imin \gets \lceil \log_k \wmin \rceil$ and $\imax \gets \log_k \wmax$.\\
			Create new instances of the unweighted greedy algorithm named $\Greedy_{\imin},\allowbreak \Greedy_{\imin + 1},\allowbreak \dotsc, \Greedy_{\imax}$.
		}
		\Else
		{
			Update $\wmax \gets \max\{\wmax, w(u)\}$ and $\imax \gets \log_k \wmax$. If the value of $\wmax$ increased following this update, create new instances of unweighted greedy named $\Greedy_{\pimax+1}, \Greedy_{\pimax + 2}, \dotsc, \Greedy_{\imax}$, where $\pimax$ is the old value of $\imax$.\footnotemark\label{line:create}\\
			
			Update $\wmin \gets \wmax / (2gk)^2$ and $\imin \gets \lceil \log_k \wmin \rceil$. If the value of $\wmin$ increased following this update, delete the instances of unweighted greedy named $\Greedy_{\pimin},\allowbreak \Greedy_{\pimin + 1}, \dotsc, \Greedy_{\imin - 1}$, where $\pimin$ is the old value of $\imin$. In contrast, if the value of $\wmin$ decreased following the update, copy $\Greedy_{\pimin}$ into new instances of unweighted greedy named $\Greedy_{\imin},\allowbreak \Greedy_{\imin + 1},\allowbreak \dotsc, \Greedy_{\pimin - 1}$.\label{line:delete}
		}

		\If{$w(u) \geq \wmin$}
		{
			Let $i_u \gets \log_k w(u)$.\\
			Feed $u$ to $\Greedy_{\imin},\Greedy_{\imin + 1},\dotsc,\Greedy_{i_u}$.
		}
	}
		
	\BlankLine

	Let $C_i$ denote the output of $\Greedy_i$ for every $\imin \leq i \leq \imax$. \\

	Let $T \gets \varnothing$. \\
	\For{every $\imin \leq i \leq \imax$ in descending order}
	{
		Greedily add elements from $C_i$ to $T$ as long as this is possible. \\
		Let $T_i$ denote the current value of $T$.
	}
	\Return{$T$}.

\end{algorithm}

We now get to the analysis of \refalg{algorithm:general_weights}, and let us begin by bounding its space complexity. Let $g(h)$, $\imin(h)$, $\imax(h)$, $\wmin(h)$ and $\wmax(h)$ denote the values of $g$, $\imin$, $\imax$, $\wmin$ and $\wmax$, respectively, at the end of iteration number $h$ of Algorithm~\ref{algorithm:general_weights}.

\begin{lemma} \label{lem:app_space}
\refalg{algorithm:general_weights}  can be implemented using a space complexity of $O(\rho (\log k + \log \rho) /\log k)$.
\end{lemma}

\begin{proof}
\footnotetext{As written, Line~\ref{line:create} might create a large number of instances of unweighted greedy when there is a large increase in $\wmax$. However, when this happens most of the newly created instances are immediately deleted by Line~\ref{line:delete}. A smart implementation of Algorithm~\ref{algorithm:general_weights} can avoid the creation of unweighted greedy instances that are destined for such immediate deletion, and this is crucial for the analysis of the space complexity of Algorithm~\ref{algorithm:general_weights} in the proof of Lemma~\ref{lem:app_space}.}
Using the same argument used in the proof of Lemma~\ref{lem:space_complexity}, it can be shown that the space complexity of Algorithm~\ref{algorithm:general_weights} is upper bounded by $O(\rho)$ times the maximum number of unweighted greedy instances maintained by the algorithm at the same time. By making the deletions of unweighted greedy instances precede the creation of new instances within every given iteration of the main loop of Algorithm~\ref{algorithm:general_weights} (and avoiding the creation of instances that need to be immediately deleted), it can be guaranteed that the maximum number of instances of unweighted greedy maintained by Algorithm~\ref{algorithm:general_weights} at any given time is exactly $\max_{1 \leq h \leq n} \{\imax(h) - \imin(h) + 2\}$. Thus, the algorithm's space complexity is at most
\begin{align*}
	&
	O(\rho) \cdot \max_{1 \leq h \leq n} \{\imax(h) - \imin(h) + 2\}
	=
	O(\rho) \cdot \max_{1 \leq h \leq n} \{\log_k \wmax(h) - \log_k \lceil \wmin(h) \rceil + 2\}\\
	\leq{} &
	O(\rho) \cdot \max_{1 \leq h \leq n} \left\{\log_k \left(\frac{\wmax}{\wmin} \right) + 2\right\}
	=
	O(\rho) \cdot \max_{1 \leq h \leq n} \left\{\log_k (2k \cdot g(h))^2 + 2\right\}\\
	\leq {} &
	O(\rho) \cdot [\log_k(2 \rho k)^2 + 2]
	\leq 
	O(\rho) \cdot \frac{2 \ln \rho + 4\ln k + 2}{\ln k}
	\enspace , 
\end{align*}
where the second inequality is due to the fact that $g$ is always the size of an independent set, and thus, cannot exceed $\rho$.
\end{proof}

Our next objective is to analyze the approximation ratio of Algorithm~\ref{algorithm:general_weights}. Like in the toy analysis presented above for the case in which the algorithm has access to $\rho$ and $\max_{u \in \cN} w(u)$, the analysis we present starts by upper bounding the total weight of the discarded elements. However, to do that we need the following technical observation, which can be proved by induction.
\begin{observation} \label{obs:nice}
Algorithm~\ref{algorithm:general_weights} maintains the invariant that, at the end of every one of its loops, if an element $u \in \cN$ was fed to some instance of unweighted greedy currently kept by the algorithm, then it was fed exactly to those instances associated with a weight of at most $\log_k w(u)$.
\end{observation}

We say that an element $u \in \cN$ is \emph{discarded} by Algorithm~\ref{algorithm:general_weights} if $u$ was never fed to the final instance $\Greedy_{\imin(n)}$ (during the execution of Algorithm~\ref{algorithm:general_weights} there might be multiple instances of unweighted greedy named $\Greedy_i$ for $i = \imin(n)$---by \emph{final instance} we mean the last of these instances). Let $F$ be the set of discarded elements.

\begin{lemma}
$w(OPT \cap F) \leq \frac{1}{2} \cdot w(OPT)$.
\end{lemma}
\begin{proof}
For every $1 \leq i \leq |OPT \cap F|$, let $u_i$ be the $i$-th element of $OPT \cap F$ to arrive, and let $h_i$ be its location in the input stream. Given Observation~\ref{obs:nice}, the fact that $u_i \in F$ implies that $u_i$ was not feed to the final instance $\Greedy_{\log_k w(u)}$, which can only happen if an instance named $\Greedy_{\log_k w(u)}$ either did not exist when $u_i$ arrived or was deleted at some point after $u_i$'s arrival. Thus, $\imin(h'_i) > \log_k w(u_i)$ for some $h_i \leq h'_i \leq n$.

The crucial observation now is that $g(h'_i) \geq g(h_i) \geq i/k$ because by the time $u_i$ arrives there are already $i$ elements of $OPT$ that arrived, and these elements form together an independent set of size $i$ (recall that $g$ is a $k$-approximation for the maximum size of an independent set consisting only of elements that already arrived). Thus, we get
\[
	w(u_i)
	=
	2^{\log_k w(u_i)}
	\leq
	2^{\imin(h'_i) - 1}
	\leq
	\wmin(h'_i)
	=
	\frac{\wmax(h'_i)}{(2k \cdot g(h'_i))^2}
	\leq
	\frac{\max_{u \in \cN} w(u)}{(2k \cdot (i/k))^2}
	\leq
	\frac{w(OPT)}{4i^2}
	\enspace,
\]
where the first inequality holds since $\imin(h'_i) > \log_k w(u_i)$ and both $\imin(h'_i)$ and $\log_k w(u_i)$ are integers. Adding up the last inequality over $1 \leq i \leq |OPT \cap F|$ yields
\[
	w(OPT \cap F) 
	=
	\sum_{i=1}^{|OPT \cap F|} \mspace{-9mu} w(u_i)
	\leq
	\sum_{i=1}^{|OPT \cap F|}{\frac{w(OPT)}{4 i^2}}
	\leq
	\frac{w(OPT)}{4} \cdot \left[1 + \int_1^{\infty}{i^{-2}}\right]
	=
	\frac{w(OPT)}{2}
	\enspace .
	\qedhere
\]
\end{proof}

The next lemma shows that Algorithm~\ref{algorithm:general_weights} has a good approximation ratio with respect to the non-discarded elements of $OPT$.

\begin{lemma} \label{lemma}
$w(OPT \setminus F) \leq 2k \cdot w(T)$.
\end{lemma}

\begin{proof}
Observe that $(\cN \setminus F, \cI \cap 2^{\cN \setminus F})$ is a $k$-extendible system, derived from $(\cN, \cI)$ by removing all elements of $F$.
In addition, all the weights of the elements of this set system are powers of $k$, and thus, by Proposition~\ref{prop:k_approx_for_k_power}, Algorithm~\ref{cs_alg} achieves $2k$-approximation for the problem of finding a maximum weight independent set of $(\cN \setminus F, \cI \cap 2^{\cN \setminus F})$. In other words, when Algorithm~\ref{cs_alg} is fed only the elements of $\cN \setminus F$, its output set $T'$ obeys $w(OPT') \leq 2k \cdot w(T')$, where $OPT'$ is an arbitrary maximum weight set independent set of $(\cN \setminus F, \cI \cap 2^{\cN \setminus F})$. 

We now note that one consequence of Observation~\ref{obs:nice} is that, by the time Algorithm~\ref{algorithm:general_weights} terminates, the instances $\Greedy_{\imin(n)}, \Greedy_{\imin(n) + 1}, \dotsc, \Greedy_{\imax(n)}$ it maintains receive exactly the input received by the corresponding instances in Algorithm~\ref{cs_alg} when the last algorithm gets only the elements of $\cN \setminus F$ as input. Since Algorithms~\ref{cs_alg} and~\ref{algorithm:general_weights} compute their outputs based on the outputs of $\Greedy_{\imin(n)}, \Greedy_{\imin(n) + 1}, \dotsc, \Greedy_{\imax(n)}$ in the same way, this implies that the output set $T$ of Algorithm~\ref{algorithm:general_weights} is identical to the output set $T'$ produced by Algorithm~\ref{cs_alg} when this algorithm is given only the elements of $\cN \setminus F$ as input.

Combining the above observations, we get
\[
	w(T)
	=
	w(T')
	\geq
	\frac{w(OPT')}{2k}
	\geq
	\frac{w(OPT \setminus F)}{2k}
	\enspace,
\]
where the last inequality holds since $OPT'$ is a maximum weight independent set in $(\cN \setminus F, \cI \cap 2^{\cN \setminus F})$ and $OPT \setminus F$ is independent in this set system. The lemma now follows by rearranging the last inequality.
\end{proof}

\begin{corollary} \label{cor:app_approximation}
$w(OPT) \leq 4k \cdot w(T)$, and thus, the approximation ratio of Algorithm~\ref{algorithm:general_weights} is at most $4k$.
\end{corollary}
\begin{proof}
Combining the last two lemmata, one gets 
\[
	\frac{w(OPT)}{2}
	\leq
	w(OPT) - w(OPT \cap F)
	=
	w(OPT \setminus F)
	\leq
	2k \cdot w(T)
	\enspace.
\]
The corollary now follows by rearranging the above inequality.
\end{proof}

We conclude the section by noticing that Proposition~\ref{prop:k_approx_for_k_power_general} is an immediate consequence of Lemma~\ref{lem:app_space} and Corollary~\ref{cor:app_approximation}.

\end{document}